\documentclass[11pt,a4paper]{article}
\pdfoutput=1 
\usepackage[utf8]{inputenc}
\usepackage{graphicx}

\usepackage[dvipsnames]{xcolor}

\usepackage{float}
\usepackage{cite}
                                                                                        
\usepackage{ifthen, mciteplus} % for conditional statements

\usepackage{slashbox}

\usepackage{url}
\newboolean{pdflatex}
\setboolean{pdflatex}{true} % False for eps figures         

\newboolean{articletitles}
\setboolean{articletitles}{true} % False removes titles in references                                                               

\newboolean{uprightparticles}
\setboolean{uprightparticles}{false} %True for upright particle symbols          
\newboolean{inbibliography}
\setboolean{inbibliography}{true} %True once you enter the bibliography

\usepackage{xspace} 
\usepackage{upgreek}
\usepackage{amsmath}

\def\lhcb   {\mbox{LHCb}\xspace}

\def\lhc    {\mbox{LHC}\xspace}

\def\MagUp {\mbox{\em Mag\kern -0.05em Up}\xspace}

\ifthenelse{\boolean{uprightparticles}}%
{

 \def\Pmu         {\ensuremath{\upmu}\xspace}

 \def\Ppsi        {\ensuremath{\uppsi}\xspace}

 \def\PDelta      {\ensuremath{\Delta}\xspace}                 
 \def\PXi         {\ensuremath{\Xi}\xspace}                 
 \def\PLambda     {\ensuremath{\Lambda}\xspace}                 
 \def\PSigma      {\ensuremath{\Sigma}\xspace}                 
 \def\POmega      {\ensuremath{\Omega}\xspace}                 
 \def\PUpsilon    {\ensuremath{\Upsilon}\xspace}

 \def\PB      {\ensuremath{\mathrm{B}}\xspace}                 
                  
 \def\PD      {\ensuremath{\mathrm{D}}\xspace}

 \def\PJ      {\ensuremath{\mathrm{J}}\xspace}                 
 \def\PK      {\ensuremath{\mathrm{K}}\xspace}

 \def\Pb      {\ensuremath{\mathrm{b}}\xspace}

 \def\Pe      {\ensuremath{\mathrm{e}}\xspace}

 \def\Pi      {\ensuremath{\mathrm{i}}\xspace}

 \def\Pp      {\ensuremath{\mathrm{p}}\xspace}

 \def\Ps      {\ensuremath{\mathrm{s}}\xspace}

 \def\thebaroffset{0.0em}
}
{

 \def\Pmu         {\ensuremath{\mu}\xspace}

 \def\Ppsi        {\ensuremath{\psi}\xspace}                 
                  
 \mathchardef\PDelta="7101
 \mathchardef\PXi="7104
 \mathchardef\PLambda="7103
 \mathchardef\PSigma="7106
 \mathchardef\POmega="710A
 \mathchardef\PUpsilon="7107
                  
 \def\PB      {\ensuremath{B}\xspace}                 
                  
 \def\PD      {\ensuremath{D}\xspace}

 \def\PJ      {\ensuremath{J}\xspace}                 
 \def\PK      {\ensuremath{K}\xspace}

 \def\Pb      {\ensuremath{b}\xspace}

 \def\Pe      {\ensuremath{e}\xspace}

 \def\Pi      {\ensuremath{i}\xspace}

 \def\Pp      {\ensuremath{p}\xspace}

 \def\Ps      {\ensuremath{s}\xspace}

 \def\thebaroffset{0.18em}
}
\newcommand{\offsetoverline}[2][\thebaroffset]{\kern #1\overline{\kern -#1 #2}}%

\makeatletter
\ifcase \@ptsize \relax% 10pt
  \newcommand{\miniscule}{\@setfontsize\miniscule{4}{5}}% \tiny: 5/6
\or% 11pt
  \newcommand{\miniscule}{\@setfontsize\miniscule{5}{6}}% \tiny: 6/7
\or% 12pt
  \newcommand{\miniscule}{\@setfontsize\miniscule{5}{6}}% \tiny: 6/7
\fi
\makeatother

\DeclareRobustCommand{\optbar}[1]{\shortstack{{\miniscule (\rule[.5ex]{1.25em}{.18mm})}
  \\ [-.7ex] $#1$}}

\def\epem       {{\ensuremath{\Pe^+\Pe^-}}\xspace}
\def\mumu       {{\ensuremath{\Pmu^+\Pmu^-}}\xspace}

\def\ellm       {{\ensuremath{\ell^-}}\xspace}
\def\ellp       {{\ensuremath{\ell^+}}\xspace}

\def\squark    {{\ensuremath{\Ps}}\xspace}

\def\bquark    {{\ensuremath{\Pb}}\xspace}

\def\kaon    {{\ensuremath{\PK}}\xspace}
\def\KorKbar {\kern \thebaroffset\optbar{\kern -\thebaroffset \PK}{}\xspace}

\def\Kstarz  {{\ensuremath{\kaon^{*0}}}\xspace}

\def\DorDbar {\kern \thebaroffset\optbar{\kern -\thebaroffset \PD}\xspace}

\def\B       {{\ensuremath{\PB}}\xspace}

\def\BorBbar {\kern \thebaroffset\optbar{\kern -\thebaroffset \PB}\xspace}

\def\Bd      {{\ensuremath{\B^0}}\xspace}

\def\BdorBdbar {\kern \thebaroffset\optbar{\kern -\thebaroffset \Bd}\xspace}

\def\Bs      {{\ensuremath{\B^0_\squark}}\xspace}

\def\BsorBsbar {\kern \thebaroffset\optbar{\kern -\thebaroffset \Bs}\xspace}

\def\jpsi     {{\ensuremath{{\PJ\mskip -3mu/\mskip -2mu\Ppsi\mskip 2mu}}}\xspace}

\def\Y#1S{\ensuremath{\PUpsilon{(#1S)}}\xspace}

\def\proton      {{\ensuremath{\Pp}}\xspace}

\def\Lz          {{\ensuremath{\PLambda}}\xspace}

\def\LorLbar     {\kern \thebaroffset\optbar{\kern -\thebaroffset \PLambda}\xspace}

\def\Lb           {{\ensuremath{\Lz^0_\bquark}}\xspace}

\def\BF         {{\ensuremath{\mathcal{B}}}\xspace}

\newcommand{\decay}[2]{\ensuremath{#1\!\to #2}\xspace} 

\def\to                 {\ensuremath{\rightarrow}\xspace}

\def\qsq       {{\ensuremath{q^2}}\xspace}

\def\BdToKstmm    {\decay{\Bd}{\Kstarz\mup\mun}}

\def\AT#1     {\ensuremath{A_{\mathrm{T}}^{#1}}\xspace}           % 2

\def\ctl       {\ensuremath{\cos{\theta_\ell}}\xspace}

\def\C#1      {\ensuremath{\mathcal{C}_{#1}}\xspace}                       % 9
\def\Cp#1     {\ensuremath{\mathcal{C}_{#1}^{'}}\xspace}                    % 7
\def\Ceff#1   {\ensuremath{\mathcal{C}_{#1}^{\mathrm{(eff)}}}\xspace}        % 9  
\def\Cpeff#1  {\ensuremath{\mathcal{C}_{#1}^{'\mathrm{(eff)}}}\xspace}       % 7
\def\Ope#1    {\ensuremath{\mathcal{O}_{#1}}\xspace}                       % 2
\def\Opep#1   {\ensuremath{\mathcal{O}_{#1}^{'}}\xspace}                    % 7

\newcommand{\aunit}[1]{\ensuremath{\text{\,#1}}}       
\newcommand{\tev}{\aunit{Te\kern -0.1em V}\xspace}
\newcommand{\gev}{\aunit{Ge\kern -0.1em V}\xspace}
\newcommand{\mev}{\aunit{Me\kern -0.1em V}\xspace}
\newcommand{\kev}{\aunit{ke\kern -0.1em V}\xspace}
\newcommand{\ev}{\aunit{e\kern -0.1em V}\xspace}
\newcommand{\mevc}{\ensuremath{\aunit{Me\kern -0.1em V\!/}c}\xspace}
\newcommand{\gevc}{\ensuremath{\aunit{Ge\kern -0.1em V\!/}c}\xspace}
\newcommand{\mevcc}{\ensuremath{\aunit{Me\kern -0.1em V\!/}c^2}\xspace}
\newcommand{\gevcc}{\ensuremath{\aunit{Ge\kern -0.1em V\!/}c^2}\xspace}
\newcommand{\gevgevcccc}{\ensuremath{\gev^2\!/c^4}\xspace} % for q^2

\def\fb   {\ensuremath{\aunit{fb}}\xspace}
\def\invfb   {\ensuremath{\fb^{-1}}\xspace}

\def\gsim{{~\raise.15em\hbox{$>$}\kern-.85em
          \lower.35em\hbox{$\sim$}~}\xspace}
\def\lsim{{~\raise.15em\hbox{$<$}\kern-.85em
          \lower.35em\hbox{$\sim$}~}\xspace}

\def\pt         {\ensuremath{p_{\mathrm{T}}}\xspace}

\def\tell1  {TELL1\xspace}
\def\ukl1   {UKL1\xspace}

\def\lplm{\ensuremath{\ellp\ellm}\xspace}

\def\Lst{\ensuremath{\PLambda^*}\xspace}

\def\BdToKstll{\decay{\Bd}{\Kstarz \lplm}}
\def\BdToKstmm{\decay{\Bd}{\Kstarz \mumu}}

\def\LbTopKll{\decay{\Lb}{\proton\kaon \lplm}}

\def\LbTopKmm{\decay{\Lb}{\proton\kaon \mumu}}
\def\LbTopKee{\decay{\Lb}{\proton\kaon \epem}}

\def\LbTopKJPsi{\decay{\Lb}{\proton\kaon \jpsi}}

\def \ctl {\ensuremath{\cos{\theta_\ell}}\xspace}

\def \AFB {\ensuremath{A_{FB}^\ell}\xspace}
\def \Sonecc{\ensuremath{S_{1cc}}\xspace}
\def \dG{\ensuremath{d\Gamma/dq^2/\Gamma_{\Lb}}\xspace}

\def\LbToLstll{\decay{\Lb}{\Lz(1520) \lplm}}
\def\LbToLstmm{\decay{\Lb}{\Lz(1520) \mumu}}
\def\btosll{\decay{\bquark}{\squark\lplm}}
\def\BF         {{\ensuremath{\mathcal{B}}}\xspace}

\title{\bf Prospects for New Physics searches with  \LbToLstll decays }
\author{Yasmine Amhis$^1$,  S\'ebastien Descotes-Genon$^1$,  Carla Marin Benito$^{1*}$,\\ Mart\'in Novoa-Brunet$^1$\footnote{Corresponding Authors},  Marie-H\'el\`ene Schune$^1$  \\
 }

\begin{document}
\date{}

\maketitle

\begin{center}{\footnotesize \it
\noindent
$^{1}$ Universit\'e Paris-Saclay, CNRS/IN2P3, IJCLab, 91405 Orsay, France}

\vspace{0.5cm}

{\footnotesize
{{Email:~}}{\bf\color{blue} yasmine.amhis@ijclab.in2p3.fr,\\ sebastien.descotes-genon@ijclab.in2p3.fr,\\ carla.marin@ijclab.in2p3.fr,\\ martin.novoa@ijclab.in2p3.fr,\\ marie-helene.schune@ijclab.in2p3.fr \\}
}
\end{center}
\bigskip

\hrule
\begin{abstract}\noindent
We present the prospects of an angular analysis of the \LbToLstll decay. Using the expected yield in the current dataset
collected at the LHCb experiment, as well as the foreseen ones after the LHCb upgrades, sensitivity studies are presented to determine the experimental precision on angular observables related to the lepton distribution and their potential to identify New Physics. The forward-backward lepton asymmetry at low dilepton invariant mass is particularly promising. NP scenarios favoured by the current anomalies in $b\to s\ell^+\ell^-$ decays can be distinguished from the SM case with the data collected between the Run 3 and the Upgrade 2 of the LHCb experiment.
\end{abstract}
\hrule

\section{Introduction}
\label{sec:introduction}

Over the last few years, the rare \btosll decays have shown a growing set of deviations with respect to Standard Model (SM) expectations. On one hand, there have been deviations observed in the branching ratios for $B\to K\mu^+\mu^-$~\cite{Aaij:2014pli}, $B\to K^*\mu^+\mu^-$~\cite{Aaij:2014pli,Aaij:2013iag,Aaij:2016flj}, $B_s\to \phi\mu^+\mu^-$~\cite{Aaij:2015esa} as well as for the optimised angular observables~\cite{Matias:2012xw,Descotes-Genon:2013vna} in $B\to K^*\mu^+\mu^-$~\cite{Aaij:2013aln,Aaij:2015oid,Abdesselam:2016llu,ATLAS:2017dlm,CMS:2017ivg}, with deviations up to 2.6 $\sigma$. These deviations have been recently confirmed by the analysis of part of the Run 2 data set by the LHCb collaboration~\cite{Aaij:2020nrf}.
On the other hand, no such deviations have been observed in $b\to se^+e^-$ branching ratios and angular observables, as was summarised in measurements of the $R_K$~\cite{Aaij:2014ora} and $R_{K^*}$~\cite{Aaij:2017vbb} ratios of branching ratios and $B\to K^*\ell^+\ell^-$ angular observables~\cite{Aaij:2015dea,Wehle:2016yoi} for several values of the dilepton invariant mass, hinting at a violation of lepton flavour universality (LFU).

Flavour-changing neutral currents have  been used as a powerful tool to probe quantum intermediate states much more massive than the initial and final particles. In the case of \btosll, steady theoretical progress has been achieved to understand various SM effects and define observables with smaller sensitivity to hadronic uncertainties.
Moreover, a framework for model-independent analyses in terms of the weak effective Hamiltonian (described below) has been exploited to disentangle long- and short-distance contributions to these decays. It turns out that the various deviations observed can be explained consistently and economically in terms of a few shifts in the Wilson coefficients describing short-distance physics, as could be expected from New Physics (NP) violating lepton flavour universality and coupling to muons but not (or little) to electrons~(see the updated results in Ref~\cite{Alguero:2019ptt} and other works in Refs~\cite{Descotes-Genon:2013wba,Descotes-Genon:2015uva,Capdevila:2017bsm,Aebischer:2019mlg,Ciuchini:2019usw,Alok:2019ufo,Arbey:2019duh,Kumar:2019qbv,Bhattacharya:2019dot,Biswas:2020uaq}). The corresponding violation of LFU between muons and electrons is indeed significant, around $25\%$ of the SM value for the semileptonic operator $O_{9\mu}$, with several scenarios showing an equivalent ability to explain the observed deviations~\cite{Bifani:2018zmi}.

In order to confirm these hints of NP in \btosll,
it is interesting to exploit the growing set of data from the LHCb experiment~\cite{Alves:1129809} not only to measure known observables more accurately, but also to investigate other decays probing the same physics. This is true in particular for $\Lb$ decays which offer completely different theoretical and experimental environments. A first step in this direction has been attempted through the study of the decay $\Lb\to\Lz(\to p\pi)\mu^{+}\mu^{-}$~\cite{Gutsche:2013pp,Boer:2014kda,Roy:2017dum, Das:2018iap,Das:2018sms,Blake:2017une,Bhattacharya:2019der}, showing  deviations from the SM in the branching ratio and some of the angular observables~\cite{Aaij:2015xza,Aaij:2018gwm}, but with rather large uncertainties so that these results agree both with the SM and with NP interpretations already hinted at in rare meson decays~\cite{Detmold:2016pkz,Blake:2019guk}.

Another promising possibility consists in looking at decays of the \Lb baryon into excited \Lz states through the decay chain $\Lb\to \Lz^*(\to pK)\ell^{+}\ell^{-}$. Recently, the LHCb experiment has measured the ratio $R_{\proton\kaon}$ comparing $\BF(\Lb\to pK\mu^{+}\mu^{-})$ and $\BF(\Lb\to pKe^{+}e^{-})$ for a squared dilepton invariant mass, \qsq, between 0.1 and 6 GeV$^2/c^{4}$ and a $pK$ invariant mass below 2.6 GeV/$c^{2}$~\cite{Aaij:2019bzx}. The result is compatible with SM expectations, but it suggests a suppression of $\BF(\Lb\to pK\mu^{+}\mu^{-})$ compared $\BF(\Lb\to pK e^{+}e^{-})$. However, the interpretation of this result would require a precise theoretical knowledge of the various excited $\Lz$ states contributing in this large $pK$ region (hadronic form factors, interference patterns). A deeper understanding could be achieved by focusing on a single of these resonances as an intermediate state. The most interesting one seems to be the $\Lz(1520)$ baryon, which is narrow and features prominently in previous related studies, for instance in pentaquark searches using $\Lb\to pKJ/\psi$~\cite{Aaij:2015tga}. It appears thus interesting to study both the branching ratio and the angular geometry of the decay \LbToLstll. A first theoretical study was proposed in Ref.~\cite{Descotes-Genon:2019dbw} (confirmed in Ref.~\cite{Das:2020cpv}), and we want to investigate here the prospects of studying this decay with the LHCb experiment in the near future.
The definition of the LHCb run periods and upgrades, as well as the size of the collected and expected datasets  can be found in Ref.~\cite{Aaij:2636441}.

\section{Theoretical framework}

It is possible to analyse \btosll decays using a model-independent approach, namely the effective Hamiltonian,  so that heavy/energetic degrees of freedom have been integrated out in short-distance Wilson coefficients ${\mathcal C}_{i}$, leaving only operators $O_i$ describing long-distance physics
\begin{equation}\label{eq:hameff}
{\cal H}_{\rm eff}(b\to s\ell^{+}\ell^{-})=-\frac{4G_F}{\sqrt{2}} V_{tb}V_{ts}^*\sum_i {\mathcal C}_{i}  O_i + h.c,
\end{equation}
(up to small corrections proportional to $V_{ub}V_{us}^*$ in the SM). The factorisation scale for the Wilson coefficients is taken as $\mu_b=$ 4.8 GeV. The main operators are
\begin{eqnarray}
{\mathcal{O}}_{7} &=& \frac{e}{16 \pi^2} m_b
(\bar{s} \sigma_{\mu \nu} P_R b) F^{\mu \nu} ,\qquad
{\mathcal{O}}_{{7}^\prime} = \frac{e}{16 \pi^2} m_b
(\bar{s} \sigma_{\mu \nu} P_L b) F^{\mu \nu} , \nonumber  %\label{O7}  
\\
{\mathcal{O}}_{9\ell} &=& \frac{e^2}{16 \pi^2} 
(\bar{s} \gamma_{\mu} P_L b)(\bar{\ell} \gamma^\mu \ell),\qquad\quad
{\mathcal{O}}_{{9}^\prime\ell} = \frac{e^2}{16 \pi^2} 
(\bar{s} \gamma_{\mu} P_R b)(\bar{\ell} \gamma^\mu \ell), \nonumber
\\
\label{eq:O10}
{\mathcal{O}}_{10\ell} &=&\frac{e^2}{16 \pi^2}
(\bar{s}  \gamma_{\mu} P_L b)(  \bar{\ell} \gamma^\mu \gamma_5 \ell) , \qquad
{\mathcal{O}}_{{10}^\prime\ell} =\frac{e^2}{16\pi^2}
(\bar{s}  \gamma_{\mu} P_R b)(  \bar{\ell} \gamma^\mu \gamma_5 \ell) ,
\end{eqnarray}
where $P_{L,R}=(1 \mp \gamma_5)/2$ and $m_b \equiv m_b(\mu_b)$ denotes the running $b$ quark mass in the $\overline{\mathrm{MS}}$ scheme. 
In the SM, three operators play a leading role in the discussion: the electromagnetic operator $O_7$ and the semileptonic operators $O_{9\ell}$ and $O_{10\ell}$ that differ thorugh the chirality of the emitted charged leptons. NP contributions could either modify the value of the short-distance Wilson coefficients ${\mathcal C}_{7,9,10}$, or make other operators contribute in a significant manner, such as the chirality-flipped operators ${\mathcal O}_{7',9',10'}$ defined above, or other operators (scalar, pseudoscalar, tensor).

Using this separation, one can express
the decay amplitude for
\LbToLstll in terms of Wilson coefficients and hadronic matrix elements of operators. Most of them can be expressed in terms of form factors to be computed using quark models, light-cone sum rules or lattice QCD computations. Some remaining contributions correspond to long-distance charmonium contributions which have still to be investigated for this particular decay (see below).

In Ref.~\cite{Descotes-Genon:2019dbw}, the resulting angular distribution was computed:
\begin{eqnarray}\label{eq:angobs}
&&\frac{8\pi}{3}\frac{d^4\Gamma}{dq^2d\cos{\theta_\ell}d\cos{\theta_p}d\phi}\\
&&\quad =\cos ^2\theta_p \left(L_{1c} \cos \theta_\ell+L_{1cc} \cos ^2\theta_\ell+L_{1ss} \sin ^2\theta_\ell\right)\nonumber\\
&&\qquad + \sin ^2\theta_p \left(L_{2c} \cos
   \theta_\ell+L_{2cc} \cos ^2\theta_\ell+L_{2ss} \sin ^2\theta_\ell\right)\nonumber\\
&&\qquad + \sin ^2\theta_p \left(L_{3ss} \sin ^2\theta_\ell \cos^2
   \phi+L_{4ss} \sin ^2\theta_\ell \sin \phi \cos
   \phi\right)\nonumber\\
&&\qquad +\sin \theta_p \cos \theta_p \cos \phi (L_{5s} \sin \theta_\ell+L_{5sc} \sin \theta_\ell \cos \theta_\ell)\nonumber\\
&&\qquad +\sin \theta_p \cos \theta_p\sin \phi (L_{6s} \sin
   \theta_\ell+L_{6sc} \sin \theta_\ell \cos \theta_\ell),\nonumber
\end{eqnarray}
where $\theta_p, \theta_\ell, \phi$ describe the kinematics of the decay in agreement with the kinematics of Refs.~\cite{Blake:2017une,Aaij:2015xza,Aaij:2018gwm} (up to the identifications $\theta_p=\theta_{\Lst}=\theta_b$ and $\phi=\chi$). Each angular coefficient corresponds to a sum of interferences between pairs of helicity amplitudes (they are not linearly independent:
$L_{2ss} = L_{1ss}/4 + L_{2cc}/2 - L_{1cc}/8-L_{3ss}/2$). Each helicity amplitude (12 in total) is defined in terms of the helicities of the hadrons involved and expressed in terms of
Wilson coefficients and form factors (14 in total). The expressions including the lepton mass can be found in Ref.~\cite{Das:2020cpv}.

One can define derived observables using a particular weight $\omega$ to integrate the differential decay rate over the whole phase space
\begin{equation}
X[\omega](q^2)\equiv\int
\frac{d^4\Gamma}{dq^2d\cos{\theta_\ell}d\cos{\theta_p}d\phi}
\omega(q^2,\theta_\ell,\theta_p,\phi)
d\cos{\theta_\ell}d\cos{\theta_p}d\phi.
\end{equation}

The differential decay width is
\begin{equation}
\frac{d\Gamma}{dq^2}=X[1]=\frac{1}{3}[L_{1cc}+2L_{1ss}+2L_{2cc}+4L_{2ss}+2L_{3ss}],
\end{equation}
which we can use to normalise the CP-averaged angular observables and the corresponding CP-asymmetries
\begin{equation}\label{eq:SandA}
S_{i}=\frac{L_i+\bar{L}_i}{d(\Gamma+\bar\Gamma)/dq^2}, \qquad A_{i}=\frac{L_i-\bar{L}_i}{d(\Gamma+\bar\Gamma)/dq^2}.
\end{equation}

One can also define various derived quantities from these angular observables, in particular the 
 forward-backward asymmetry with respect to the leptonic scattering angle normalised to the differential rate
 \begin{equation}\label{eq:AFB}
 A^\ell_{FB}=X\left[\frac{\mathrm{ sgn}[\cos\theta_\ell]}{d\Gamma/dq^2}\right]=\frac{3 (L_{1c}+2 L_{2c})}{2 (L_{1cc}+2 (L_{1ss}+L_{2cc}+2 L_{2ss}+L_{3ss}))}.
 \end{equation}
 The CP-averaged version of this asymmetry is defined by taking the ratio of the CP averages of the numerator and the denominator in
 Eq.~(\ref{eq:AFB}).
 
 In Ref.~\cite{Descotes-Genon:2019dbw}, the sensitivity of these angular observables on hadronic uncertainties and on NP contributions was investigated. Several of them were shown to exhibit sensitivity with respect to NP contributions currently favoured by global fits to the \btosll data, even though the hadronic uncertainties were certainly overestimated due to a limited knowledge of the form factors. The estimates provided in Ref.~\cite{Descotes-Genon:2019dbw} were based on a quark model~\cite{Mott:2011cx}, waiting for more accurate estimates to be provided by lattice QCD computations (interestingly, the recent results in Refs.~\cite{Meinel:2016cxo,Meinel:2020owd} show a good agreement with this quark model), and assuming that long-distance charmonium contributions were small in the regions of interest. 
 
 Since we want to discuss the potential of experimental measurements,
 we consider a more complete approach to determine theoretical uncertainties.
 On one hand, we keep the same form factors as in Ref.~\cite{Descotes-Genon:2019dbw} with various assumptions on the uncertainties based on the foreseen improvement of theoretical computations. On the other hand, 
 we include an estimate for the charmonium contribution inspired by estimates for \BdToKstll and $\decay{\Bd}{K \lplm}$ \cite{Capdevila:2017ert,Khodjamirian:2010vf,Khodjamirian:2012rm} at large recoil. 
 Their results suggest a contribution of order 10\% of ${\mathcal C}_{9}$
 with a moderate dependence on $q^2$ in the large-recoil region of interest (further discussion on this topic can be found in Refs.~\cite{Chrzaszcz:2018yza,Bobeth:2017vxj,Serra:2016ivr,Hurth:2020rzx,Arbey:2019duh,Hurth:2018kcq,Hurth:2017hxg,Chobanova:2017ghn,Hurth:2014vma,Ciuchini:2018anp,Ciuchini:2017mik,Ciuchini:2015qxb,Aebischer:2019mlg,Altmannshofer:2017fio,Altmannshofer:2015sma,Altmannshofer:2014rta, Alguero:2019ptt,Alguero:2019pjc,Alguero:2018nvb,Capdevila:2017bsm,Capdevila:2016ivx,Descotes-Genon:2015uva,Descotes-Genon:2014uoa}). 
 
Inspired by these results, we 
 include an estimate of the charmonium contribution to our decay as an additional contribution to the SM value of ${\mathcal C}_{9\ell}$
 \begin{equation}\label{eq:C9ccmodel}
 {\mathcal C}_{9\ell} ^{\rm SM}={\mathcal C}_{9\ell}^{{\cal H}_{\rm eff}} (1+\rho e^{i\phi})\,,\qquad \rho\in[0,0.1]\,,\qquad \phi \in [0,2\pi]\,,
\end{equation}
where ${\mathcal C}_{9\ell}^{{\cal H}_{\rm eff}}$ corresponds to the Wilson coefficient appearing in Eq.~(\ref{eq:hameff}).
Admittedly, this procedure is a very rough attempt at obtaining an estimate for this effect, as it includes no dependence on $q^2$ nor on the spins of the baryons and its size is only guesstimated. 
This estimate for this baryon decay lacks thus the theoretical grounds of the dedicated calculations and sophisticated phenomenological studies for the equivalent meson 
decays. An extension to the $\Lb\to \Lz^*(\to pK)\ell^{+}\ell^{-}$ decay would be highly commendable and it will be needed in the future once measurements are available and to be interpreted within an effective theory approach, but it is clearly out of the scope of the present study limited to the potential of observing this decay at the LHCb experiment and of triggering further theoretical studies.

 The large number of (poorly known) form factors could be tackled by taking the heavy-quark limit $m_b\to\infty$
either at low or large recoil of the $\Lz$ baryon (high $q^2$ or low $q^2$), leading to two distinct effective field theories called Heavy Quark Effective Theory and  Soft-Collinear Effective Theory respectively. 
 
 In both limits, some form factors (denoted collectively as $f_g^X$ in Ref.~\cite{Descotes-Genon:2019dbw}) vanish, which means that they are expected to be small in general. Neglecting these form factors means that the amplitudes denoted as $B$ in Ref.~\cite{Descotes-Genon:2019dbw} and corresponding to transitions to a $\Lst$ with helicity 3/2 vanish. Indeed the heavy-quark limit allows one to consider the angular momentum of the heavy-quark $b$ and that of the light quarks as good quantum numbers to describe the
 $\Lb$ state and its transitions. 
 Since the light quarks are in a spin-0 diquark state
 and the heavy quark carries a spin 1/2, 
 the \btosll transition can never yield a \Lz with a helicity 3/2 in this limit~\cite{Mannel:1990vg,Feldmann:2011xf}. This assumption simplifies quite a lot the angular distribution
 \begin{eqnarray}\label{eq:angobssimpl}
&&\frac{8\pi}{3}\frac{d^4\Gamma}{dq^2d\cos{\theta_\ell}d\cos{\theta_p}d\phi}\\
&&\qquad \simeq \frac{1}{4}(1+3\cos ^2\theta_p) \left(L_{1c} \cos \theta_\ell+L_{1cc} \cos ^2\theta_\ell+L_{1ss} \sin ^2\theta_\ell\right),\nonumber
\end{eqnarray}
where all three angular observables are independent. The forward-backward asymmetry becomes:
 \begin{equation}
 \AFB\simeq \frac{3L_{1c}}{2 (L_{1cc}+2 L_{1ss})}.
 \end{equation}
In this limit, we notice that
  the angular distribution Eq.~(\ref{eq:angobssimpl}) factorises into the product of two terms, i.e. a trivial dependence on the angle describing the hadronic final state and a nontrivial dependence on the angle describing the leptonic final state.

 Interestingly, the branching ratio for \LbToLstll for $\ell=\mu$ was shown to decrease for the NP scenarious favoured to explain the deviations observed in the meson sector. This agrees perfectly with the trend shown by the recent LHCb measurement of $R_{pK}$ for $m_{pK}<2.6$ GeV$/c$ and $0.1<q^2<6$ GeV$/c^4$~\cite{Aaij:2019bzx}. Indeed this measurement involves several intermediate $\Lz$ resonances, but with a prominent contribution of the $\Lz(1520)$ baryon.
 If one assumes that this measurement is indeed dominated by the contribution of $\Lz(1520)$ and one neglects long-distance $c\bar{c}$ contributions at large $\Lz$ recoil, we can get the measured central value of 0.85 as the central value of the predictions from Ref.~\cite{Descotes-Genon:2019dbw} for the following three NP points: $C_{9\mu}^{\rm NP} = - 0.76$, or
$C_{9\mu}^{\rm NP} =-C_{10\mu}^{\rm NP}= - 0.29$,
or $C_{9\mu}^{\rm NP} =-C_{9'\mu}^{\rm NP}= - 0.99$
 (in each scenario, all the other Wilson Coefficients are purely SM). These points are in remarkable agreement with the results of global fits to \btosll and $b\to s\gamma$ transitions in $B$ meson decays~\cite{Alguero:2019ptt}. This exercise
 is obviously purely illustrative and its significance should not be overstated, but it shows the interest of identifying the fraction due to the $\Lz(1520)$ excited state in $R_{pK}$, and to measure the angular distribution of the decay $\Lb\to pK\ell^{-}\ell^{-}$ through this specific baryon intermediate state.
\section{Sensitivity studies}
\label{sec:sensitivity}
%------------------------------------    
%discussion about binning the theory
%------------------------------------
The \LbToLstll decay width and angular coefficients can be conveniently accessed experimentally 
in bins or regions of squared dilepton invariant mass, \qsq, as discussed in the introduction of this paper. Due to the available phase space in this decay,
and avoiding the region dominated by the charmonia resonances, the studies are performed in three regions: 
$[0.1, 3]$, $[3,6]$ and $[6, 8.68]$ GeV$^{2}/c^4$.
Additionally, a broader bin covering the 
central \qsq region, $[1, 6]$ GeV$^{2}/c^4$, is added to improve the experimental sensitivity in this region. 
As a first exercise to grasp the potential of an angular analysis of the \LbToLstll decay  and  due to the limited statistics available for this mode, sensitivity studies are performed 
using the simplified expression of the angular distribution presented in Eq.~(\ref{eq:angobssimpl}) and only CP-averaged observables are considered. 
We choose the CP-averaged forward-backward asymmetry in the lepton sector, \AFB, and the $S_{1cc}$ coefficient 
as the observables of interest and fix the normalisation by $\frac{1}{2} L_{1cc} + L_{1ss} = 1$.
The SM predictions for these angular observables are extracted in the \qsq bins of interest
from the computations in Ref.~\cite{Descotes-Genon:2019dbw} and are shown 
in Table~\ref{tab:theory_bins}, together with the differential decay width, \dG. Besides the conservative uncertainties used 
in Ref.~\cite{Descotes-Genon:2019dbw}, the theoretical precision obtained assuming a 
$5\%$ uncertainty on the form factors due to foreseeable improvements in lattice QCD studies~\cite{Meinel:2020owd} is also given. 
To illustrate the sensitivity of these observables to the effect of NP, 
the predictions of a scenario with a NP contribution $C_{9\mu}^{\rm NP} = - 1.11$ are also computed. Such a scenario is supported by the current global fits to $b\to s\ell^{+}\ell^{-}$ data~\cite{Alguero:2019ptt}.

\begin{table}[t]
\centering
\resizebox{\textwidth}{!}{
\begin{tabular}{|l||*{5}{c|}}\hline
\makebox[4em]{Observable} & \makebox[3em]{[0.1,3]} & \makebox[3em]{[3,6]} &
\makebox[3em]{[6,8.68]} & \makebox[3em]{[1,6]} \\\hline\hline

 \multicolumn{1}{|c||}{\dG $[10^{-9}]$} & & & & \\
 SM   & 0.397 $\pm$ 0.054 & 1.29 $\pm$ 0.18 &  3.22 $\pm$ 0.42  & 0.95 $\pm$ 0.13 \\
 SM - 5\% & 0.397 $\pm$ 0.032 & 1.29 $\pm$ 0.11 & 3.22 $\pm$ 0.28 & 0.95 $\pm$ 0.08 \\
 NP   & 0.337 $\pm$ 0.042 & 1.04 $\pm$ 0.13 & 2.58 $\pm$ 0.32 & 0.77 $\pm$ 0.10 \\
 NP - 5\% & 0.337 $\pm$ 0.023 & 1.04 $\pm$ 0.08 & 2.58 $\pm$ 0.20 & 0.77 $\pm$ 0.06 \\
  
  \hline

 \multicolumn{1}{|c||}{\AFB} & & & & \\
 SM   & 0.048 $\pm$ 0.018 & -0.127 $\pm$ 0.033 &  -0.235 $\pm$ 0.040  & -0.098 $\pm$ 0.031 \\
 SM - 5\% & 0.048 $\pm$ 0.013 & -0.127 $\pm$ 0.020 & -0.235 $\pm$ 0.022 & -0.098 $\pm$ 0.019 \\
 NP   & 0.098 $\pm$ 0.022 & -0.059 $\pm$ 0.034 & -0.166 $\pm$ 0.041 & -0.031 $\pm$ 0.032 \\
 NP - 5\% & 0.098 $\pm$ 0.016 & -0.059 $\pm$ 0.026 & -0.166 $\pm$ 0.030 & -0.031 $\pm$ 0.025 \\
  
   \hline
 
 \multicolumn{1}{|c||}{\Sonecc} & & & & \\
 SM   & 0.181 $\pm$ 0.031 &  0.242 $\pm$ 0.042  &  0.361 $\pm$ 0.051  &  0.221 $\pm$ 0.038  \\
 SM - 5\% & 0.181 $\pm$ 0.019 & 0.242 $\pm$ 0.021 & 0.361 $\pm$ 0.026 & 0.221 $\pm$ 0.020 \\
 NP   & 0.240 $\pm$ 0.038 &  0.263 $\pm$ 0.042  &  0.371 $\pm$ 0.050  &  0.246 $\pm$ 0.039  \\
 NP - 5\% & 0.240 $\pm$ 0.024 & 0.263 $\pm$ 0.022 & 0.371 $\pm$ 0.026 & 0.246 $\pm$ 0.021 \\

\hline

\end{tabular}
}
\caption{Theory predictions for \AFB and \Sonecc in the SM and in a NP model with $C_{9\mu}^{\rm NP} = - 1.11$.
The precision of the theory predictions is given both using the conservative uncertainties 
of Ref.~\cite{Descotes-Genon:2019dbw} and assuming form factor uncertainties of $5\%$. In both cases an uncertainty of 10\% is included on \C{9} to account for $c\bar c$ contributions. } 
\label{tab:theory_bins}
\end{table}

%------------------------------------    
%discussion about expected yields 
%------------------------------------
In the recent test of LFU in \LbTopKll decays by \lhcb~\cite{Aaij:2019bzx}, around 400 \LbTopKmm 
and 100 \LbTopKee signal events were observed in the \qsq region $[0.1, 6]\gevgevcccc$ and $m(pK) < 2600\mevcc$,
in a dataset corresponding to 3\invfb recorded at 7 and 8\tev and 1.7\invfb recorded at 13\tev.
The main difference between the muon and electron modes arises from the trigger and selection efficiencies
in the experimental study. In the following, we focus on the muon mode due to the larger experimental yields
but the results can be directly extrapolated to the electron case by scaling the yields accordingly.
\lhcb also published the background subtracted invariant mass of the hadronic system for \LbTopKmm candidates (available in the supplementary material of Ref.~\cite{Aaij:2019bzx}), 
from where we estimate that roughly around 90 events correspond to the \LbToLstmm decay. 
The \lhcb experiment has already recorded a total of $6\invfb$ at 13\tev and will accumulate a total of 23 and $50\invfb$
after Run 3 and Run 4 of the LHC, respectively. Moreover, it has been proposed to install an upgraded detector to 
take data during the High-Luminosity phase of the LHC to collect $300\invfb$~\cite{Aaij:2636441}.
A summary of the completed and planned \lhcb running periods is provided in Table~\ref{tab:lhcb_runs}.
Table~\ref{tab:yields} collects the estimated yields of \LbToLstmm decays in the different 
\qsq bins and running periods, extrapolated from the published \lhcb data and the SM prediction 
for the \qsq distribution. These numbers are used to estimate the sensitivity to NP of an angular 
analysis of this decay mode. 
For the electron case, fewer events are expected~\cite{Aaij:2019bzx}, although the
trigger-less readout foreseen for \lhcb from Run 3 onwards should allow a higher experimental efficiency for this mode.

\begin{table}[t]
\centering

\begin{tabular}{|r||*{4}{c|}}
\hline
Run period
&\makebox[3em]{Run 1 -- 2}&\makebox[3em]{Run 3}
&\makebox[3em]{Run 4} &\makebox[3em]{Run 5}\\\hline\hline

 Start date             & 2010          & 2022          & 2027      & 2032  \\

 End date               & 2018          & 2024          & 2030      & 2035   \\

 Center-of-mass energy  & 7, 8, 13\tev  & 13--14\tev    & 14\tev    & 14\tev \\

 Luminosity             & 9\invfb       & 23\invfb      & 50\invfb  & 300\invfb  \\

\hline       
\end{tabular}

\caption{Completed and planned \lhc runs, corresponding start and end dates, center-of-mass $pp$ collision energy and accumulated integrated luminosity expected to be recorded at \lhcb.} 
\label{tab:lhcb_runs}
\end{table}

\begin{table}[t]
\centering

\begin{tabular}{|r||*{4}{c|}}\hline
\backslashbox{\qsq bin [\gevgevcccc]}{Dataset [\invfb]}
&\makebox[3em]{9}&\makebox[3em]{23}&\makebox[3em]{50}
&\makebox[3em]{300}\\\hline\hline

 $[0.1,3]$  & 50  & 140  & 300  & 1750  \\

 $[3,6]$    & 150 & 400  & 900  & 5250  \\

 $[6,8.68]$ & 400 & 1100 & 2400 & 14000 \\

 $[1,6]$    & 190 & 510  & 1140 & 6650  \\

\hline       
\end{tabular}

\caption{Extrapolated \LbToLstmm signal yields in each \qsq bin for the accumulated luminosity 
expected at \lhcb at the end of Run 2, Run 3, Run 4 and High-Lumi LHC.} 
\label{tab:yields}
\end{table}

%------------------------------------    
%discussion about angular acceptance 
%------------------------------------
The angular distributions of the \LbToLstmm decay are distorted by the geometrical acceptance of the  detector, the trigger and the selection requirements~\cite{Aaij:2015oid}. 
The shapes of the acceptance have been estimated using a stand-alone fast simulation software called {\tt RapidSim}~\cite{Cowan:2016tnm} by applying the \lhcb geometrical acceptance and transverse momentum (\pt) requirements, as needed for the track reconstruction and background rejectio, on the final-state particles. These are known to be the dominant effects in shaping the acceptance distributions. In particular, the \pt of the muons is required to be larger than $400\mev$.
Using the simplified angular distribution of Eq.~(\ref{eq:angobs}) there is no need to model the angular acceptances of the $\phi$ and  $\cos \theta_{\proton}$ variables since they only appear as a common scale factor in the probability density function (PDF).  

% The acceptance of the $\phi$ angle is found to be flat, which is also what was observed in other analyses~\cite{LHCb-PAPER-2020-002}. On the contrary, the acceptances for $\cos \theta_{\Lst}$ and $\cos\theta_\ell$ are distorted. 
 The $\cos\theta_\ell$ acceptance curve is expected to be symmetric due to the symmetry between the two leptons in the decay with a loss of events for large $|\cos\theta_\ell|$ values due to the muon \pt requirement. This last characteristic is mainly visible in the low-\qsq region as shown in Appendix~\ref{appendix}. 
  
%    The $\cos \theta_{\Lst}$ acceptance is not symmetric due to the very different masses of the two-hadrons. This loss of symmetry implies the need of taking the acceptance into account this acceptance while defining the PDF for $\frac{d\Gamma}{d\cos{\theta_\ell}}$. 

%------------------------------------    
%discussion about the toys 
%------------------------------------

The sensitivity of the differential measurement of the \LbToLstmm decay width and angular observables to NP effects
is studied comparing the theoretical predictions for these observables in the SM and the NP scenario to the
expected experimental precision. The experimental sensitivity to the decay width is directly extracted 
from the expected signal yield in each bin given in Table~\ref{tab:yields}, assuming poisonian uncertainties on the yields and neglecting the effect of 
potential backgrounds, which are observed to be very small in this decay mode~\cite{Aaij:2019bzx}. 
One of the main experimental challenges in the selection of \LbToLstll decays is the contamination 
from other \Lst resonances that overlap in the \proton\kaon spectrum with the $\Lz(1520)$ state.
In the amplitude analysis of the related \LbTopKJPsi mode~\cite{Aaij:2015tga}
three other resonances were observed to contribute 
to the \proton\kaon mass region around the $\Lz(1520)$ state, namely $\Lz(1405)$, $\Lz(1600)$ and $\Lz(1800)$. 
However, all these resonances have spin $1/2$, in contrast to the $3/2$ spin of $\Lz(1520)$, which gives place to 
a different angular distribution that should allow to disentangle them, following a similar strategy to the one used in 
the angular analysis of \BdToKstmm to account for the $S$-wave contribution~\cite{Aaij:2020nrf}.

The experimental sensitivity to the angular observables is studied with pseudoexperiments. 
Events are generated according to a PDF that is the product of Eq.~(\ref{eq:angobssimpl}) 
and the experimental acceptance described above. Distributions of the \ctl variable are generated
using both the SM and NP predictions for the angular parameters 
and are fitted back with the same PDF, letting the \AFB and \Sonecc parameters float. 
A large number of experiments is generated for all the \qsq bins and expected signal yields 
in the different data-taking periods of LHC. The resulting distributions for the parameters, 
their uncertainties and pull distributions are examined. Small biases on the central values of 
the parameters in the low and central \qsq bins are observed in the low-statistics scenarios corresponding to the datasets expected in Run 2 and 3 of \lhcb. 
The effect is larger on \Sonecc and in the SM case but it is always 
below $20\%$ of the statistical uncertainty and can be added as a systematic to the measurement. 
The fit uncertainty is checked to provide good coverage in all the cases so 
it is taken as the experimental sensitivity to the angular observables.

\begin{table}[t]
\centering
\begin{tabular}{|l||*{4}{c|}}\hline
\makebox[6em]{Observable} 
&\makebox[3em]{9\invfb}&\makebox[3em]{23\invfb}&\makebox[3em]{50\invfb}
&\makebox[4em]{300\invfb}\\\hline\hline

 \multicolumn{1}{|c||}{${d\Gamma/dq^2/\Gamma_{\Lb}}[10^{-9}]$} &   &  &  &  \\
 ${[0.1,3]}$  & 0.060 & 0.036 & 0.024 & 0.010 \\
 ${[3,6]}$    & 0.106 & 0.064 & 0.043 & 0.018 \\
 ${[6,8.68]}$ & 0.176 & 0.107 & 0.072 & 0.030 \\
 ${[1,6]}$    & 0.070 & 0.042 & 0.029 & 0.012 \\

 \hline
 
 \multicolumn{1}{|c||}{\AFB}  &  &  &  &  \\
 ${[0.1,3]}$  & 0.140 & 0.079 & 0.053 & 0.022 \\
 ${[3,6]}$    & 0.061 & 0.037 & 0.025 & 0.010 \\
 ${[6,8.68]}$ & 0.036 & 0.022 & 0.015 & 0.006 \\
 ${[1,6]}$    & 0.058 & 0.035 & 0.023 & 0.010 \\

 \hline
 
 \multicolumn{1}{|c||}{\Sonecc} &  &  &  &  \\
 ${[0.1,3]}$  & 0.241 & 0.148 & 0.099 & 0.041 \\
 ${[3,6]}$    & 0.104 & 0.062 & 0.041 & 0.017 \\
 ${[6,8.68]}$ & 0.061 & 0.036 & 0.024 & 0.010 \\
 ${[1,6]}$    & 0.100 & 0.060 & 0.040 & 0.017 \\

 \hline       

\end{tabular}
\caption{Estimated experimental uncertainties for the differential decay width, \dG (top), \AFB (middle) and \Sonecc (bottom) for different data-taking scenarios in the considered \qsq bins in \gevgevcccc. }
\label{tab:exp_uncert}
\end{table}

\begin{figure}
    \centering
    \includegraphics[width=0.49\textwidth]{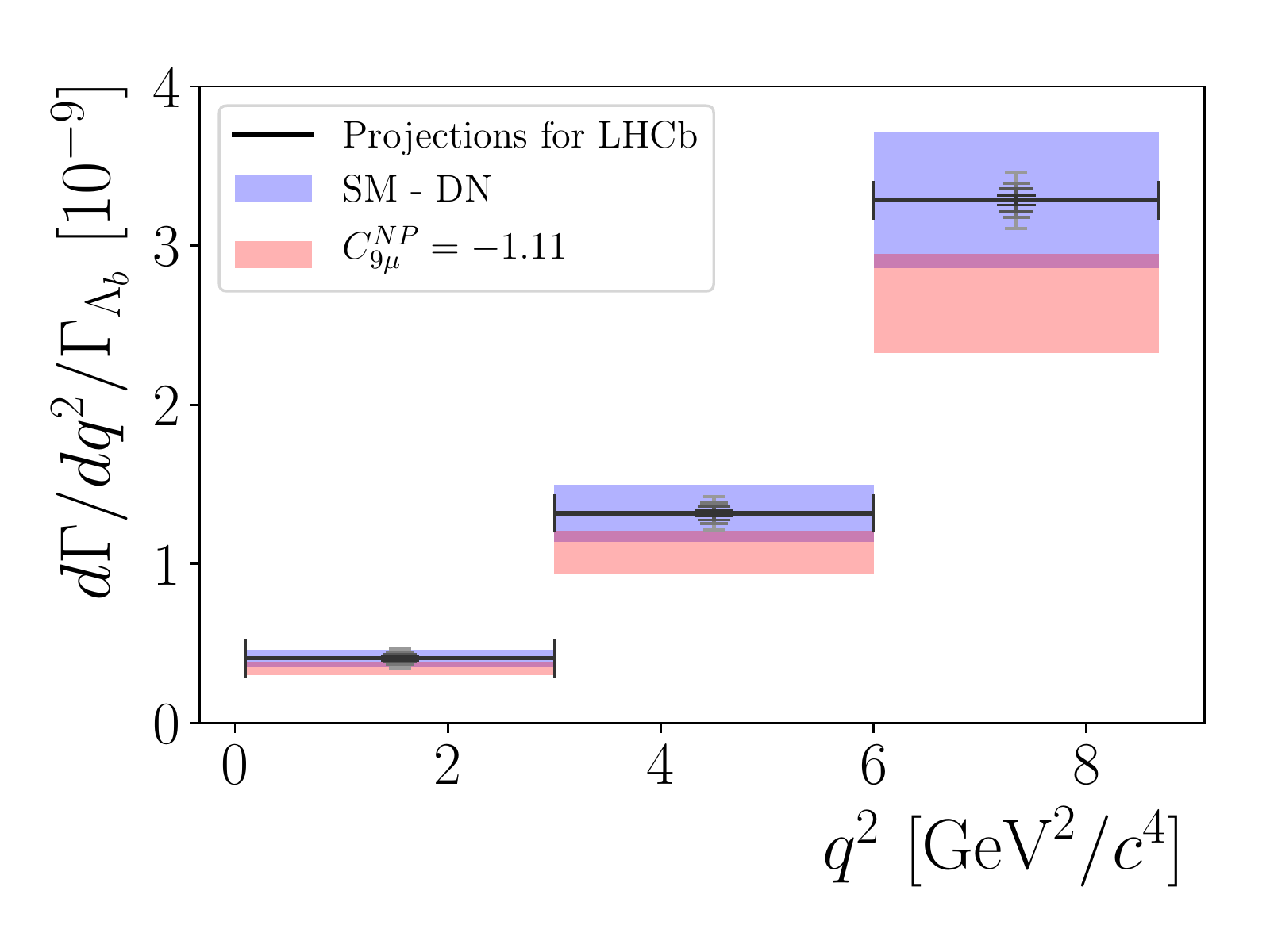}
    \includegraphics[width=0.49\textwidth]{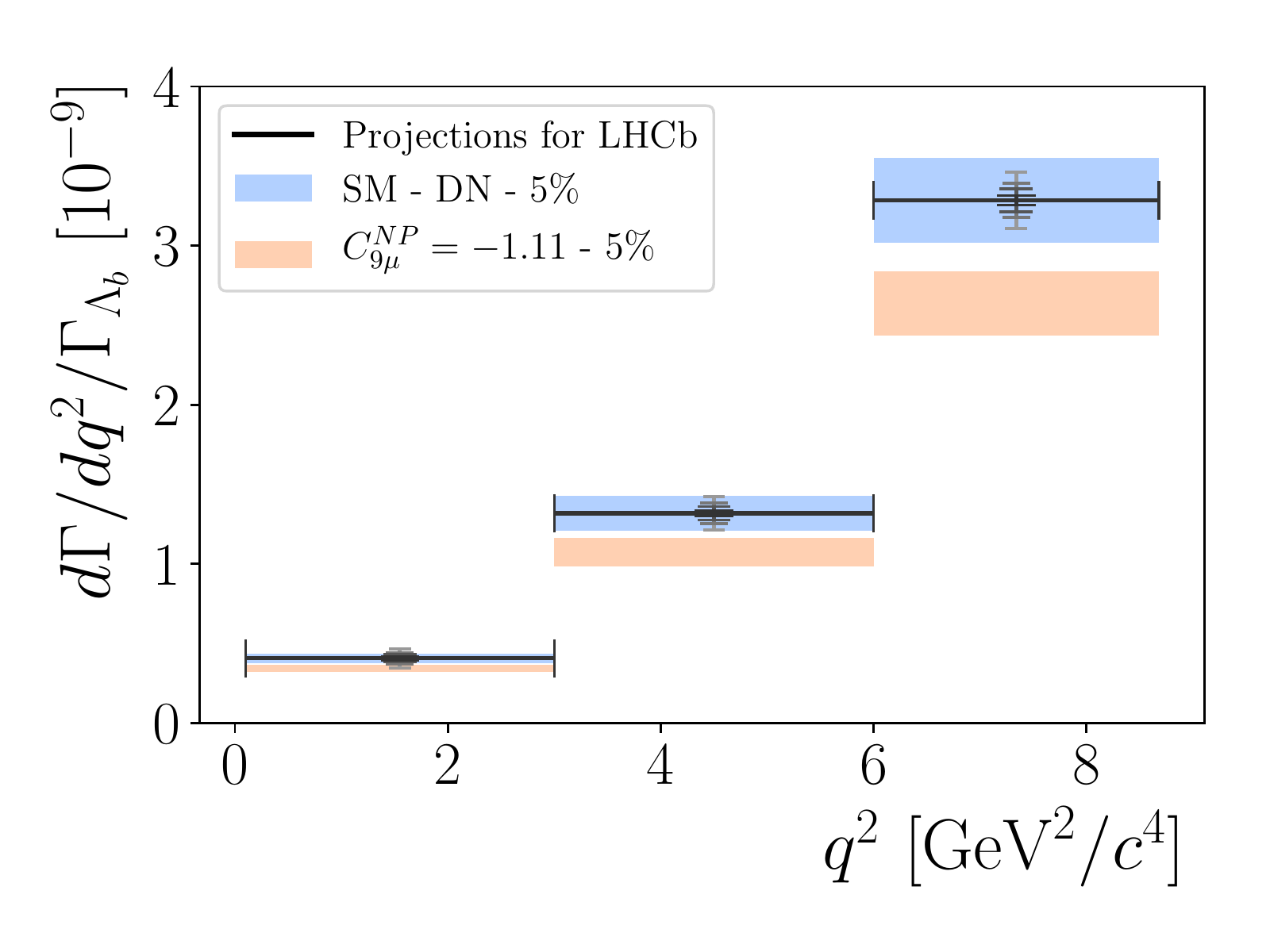} \\
    \includegraphics[width=0.49\textwidth]{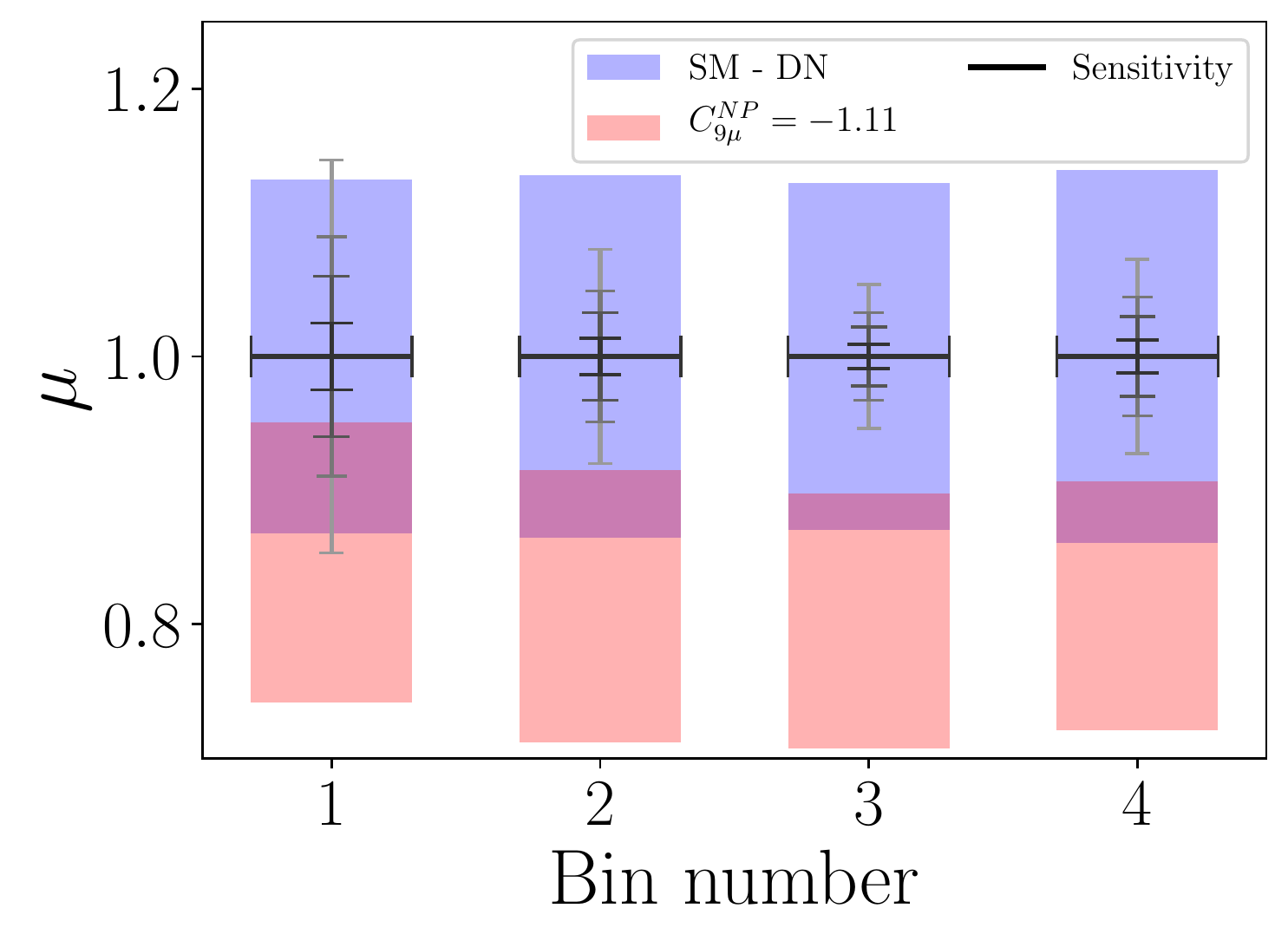}   
    \includegraphics[width=0.49\textwidth]{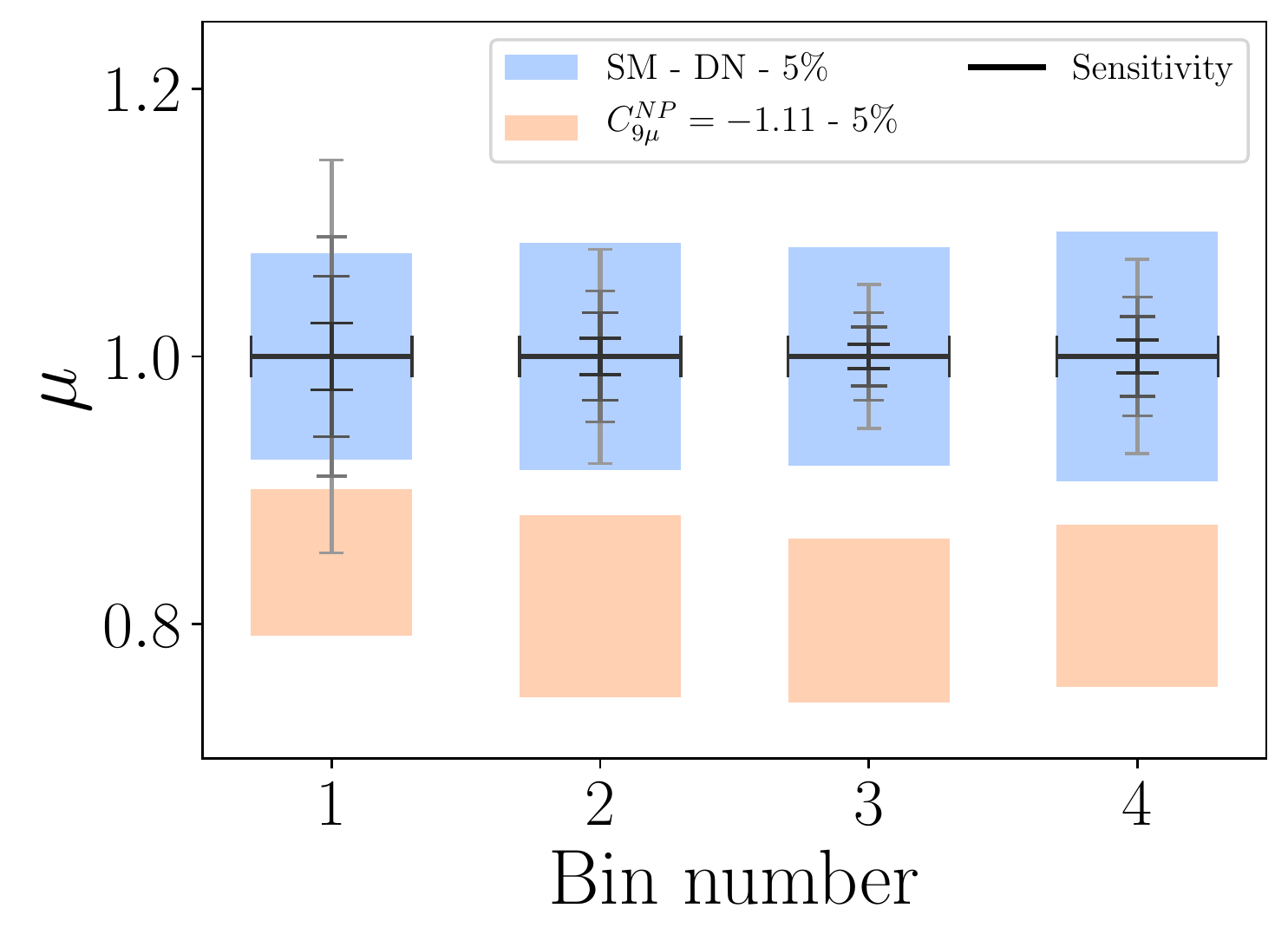}
    \caption{Theory predictions for $d\Gamma/dq^2$ in the considered \qsq bins in the SM (blue area) and the NP scenario with $C_{9\mu}^{NP}= -1.11$ (red area) from Ref.~\cite{Descotes-Genon:2019dbw}, using the nominal theory uncertainties (left) 
    and improved ones with 5\% uncertainties on the form factors (right). In both cases an uncertainty of 10\% is included on \C{9} to account for $c\bar c$ contributions. 
    The expected LHCb sensitivity with the full Run 2, Run 3, Run 4 and Upgrade 2 samples 
    is shown by grey-scale markers in increasing sensitivity.
    The bottom plots show the relative values with respect to the SM prediction in each bin ($\mu$) 
    for all the \qsq bins considered (see text for details).}
    \label{fig:sensitivity_dG}
\end{figure}

\begin{figure} [h]
    \centering
    \includegraphics[width=0.49\textwidth]{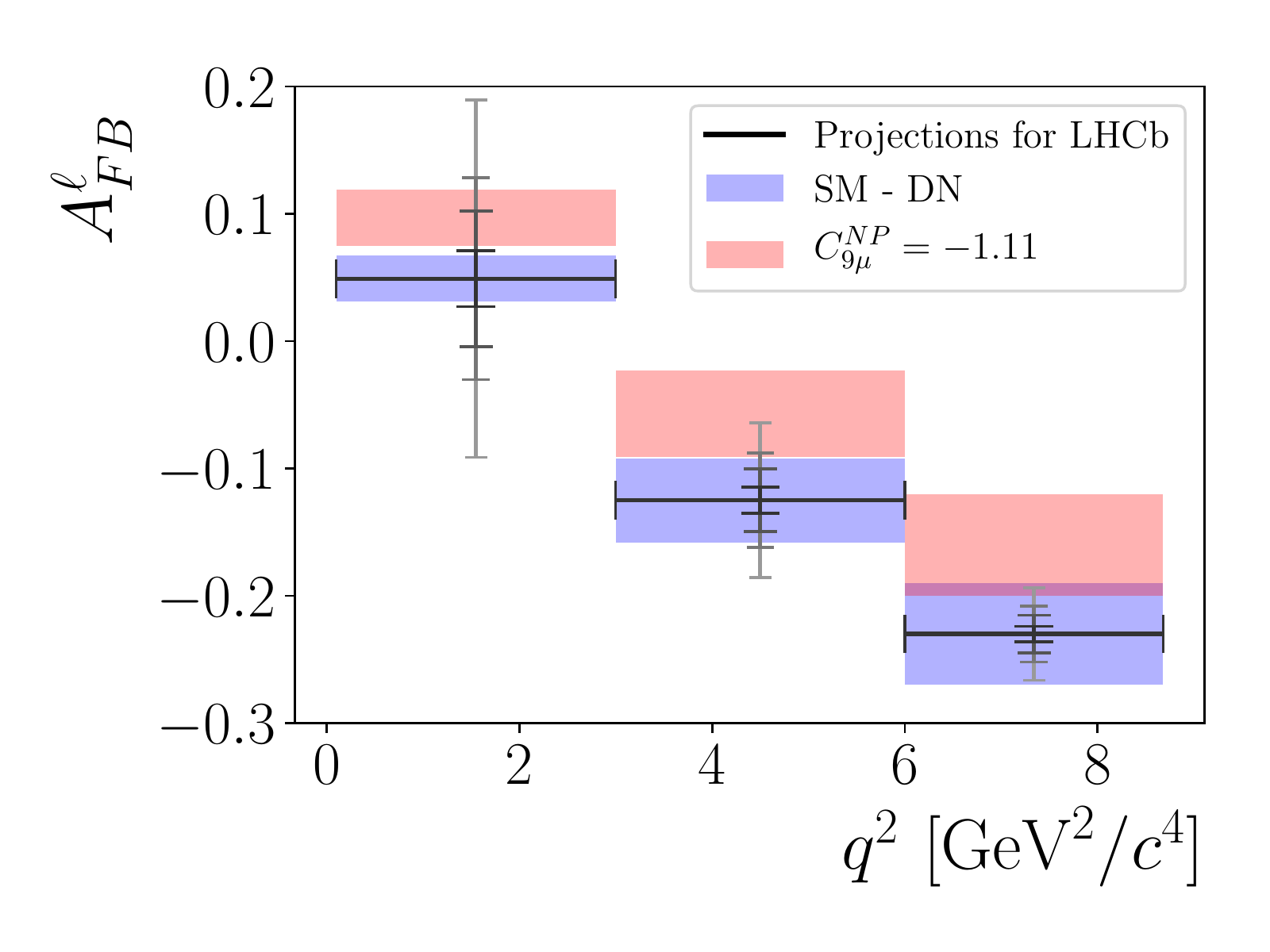}    
    \includegraphics[width=0.49\textwidth]{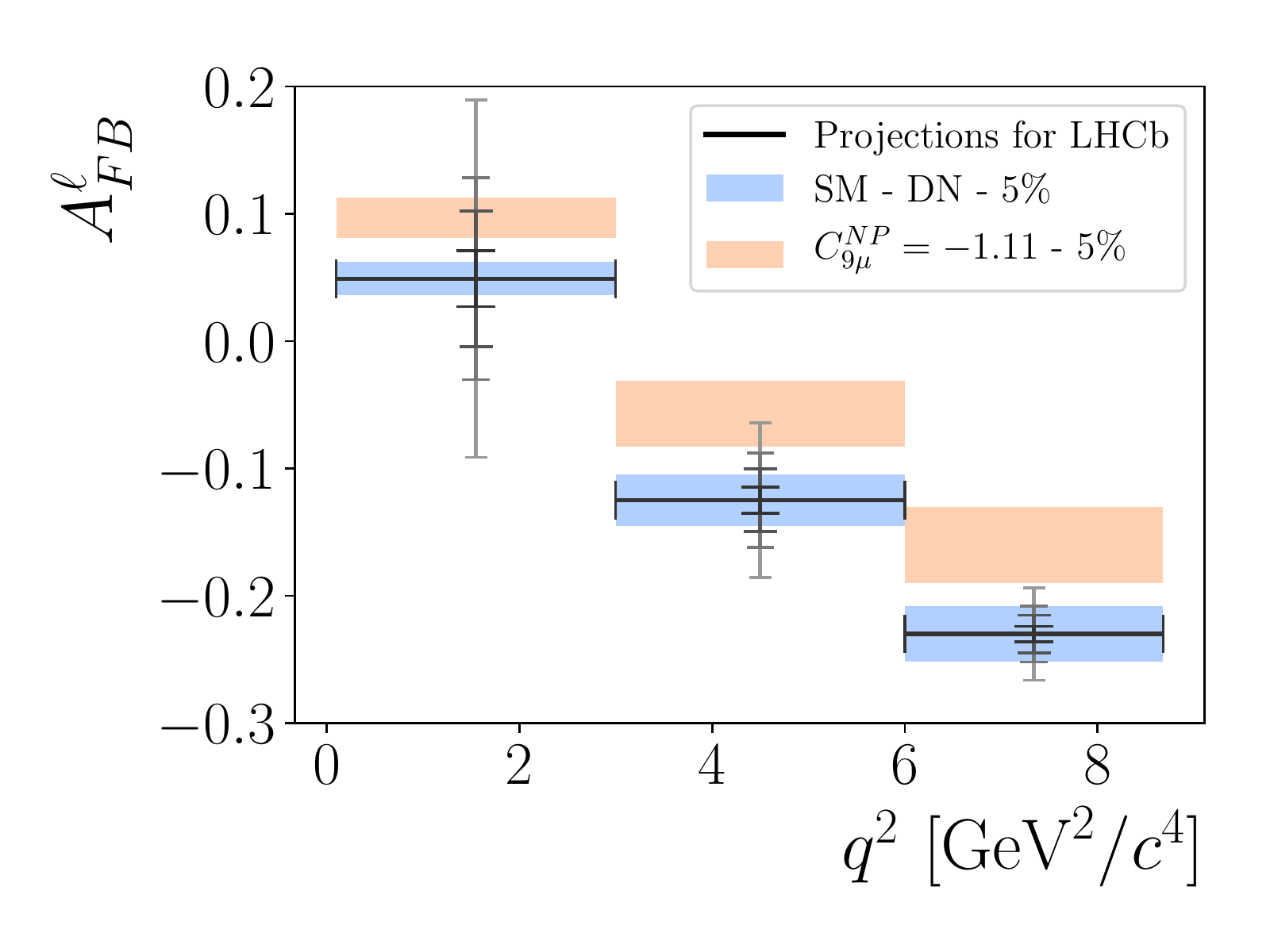}  
    \caption{Theory predictions for $A^{\ell}_{FB}$ in the considered \qsq bins in the SM (blue area) and the NP scenario with $C_{9\mu}^{\rm NP}= -1.11$ (red area) from Ref.~\cite{Descotes-Genon:2019dbw}, using the nominal theory uncertainties (left) 
    and improved ones with 5\% uncertainties on the form factors (right). In both cases an uncertainty of 10\% is included on \C{9} to account for $c\bar c$ contributions. 
    The expected LHCb sensitivity with the full Run 2, Run 3, Run 4 and Upgrade 2 samples is shown by the grey-scale markers in increasing sensitivity.}
    \label{fig:sensitivity_AFB}
\end{figure}

\begin{figure} [h]
    \centering
    \includegraphics[width=0.49\textwidth]{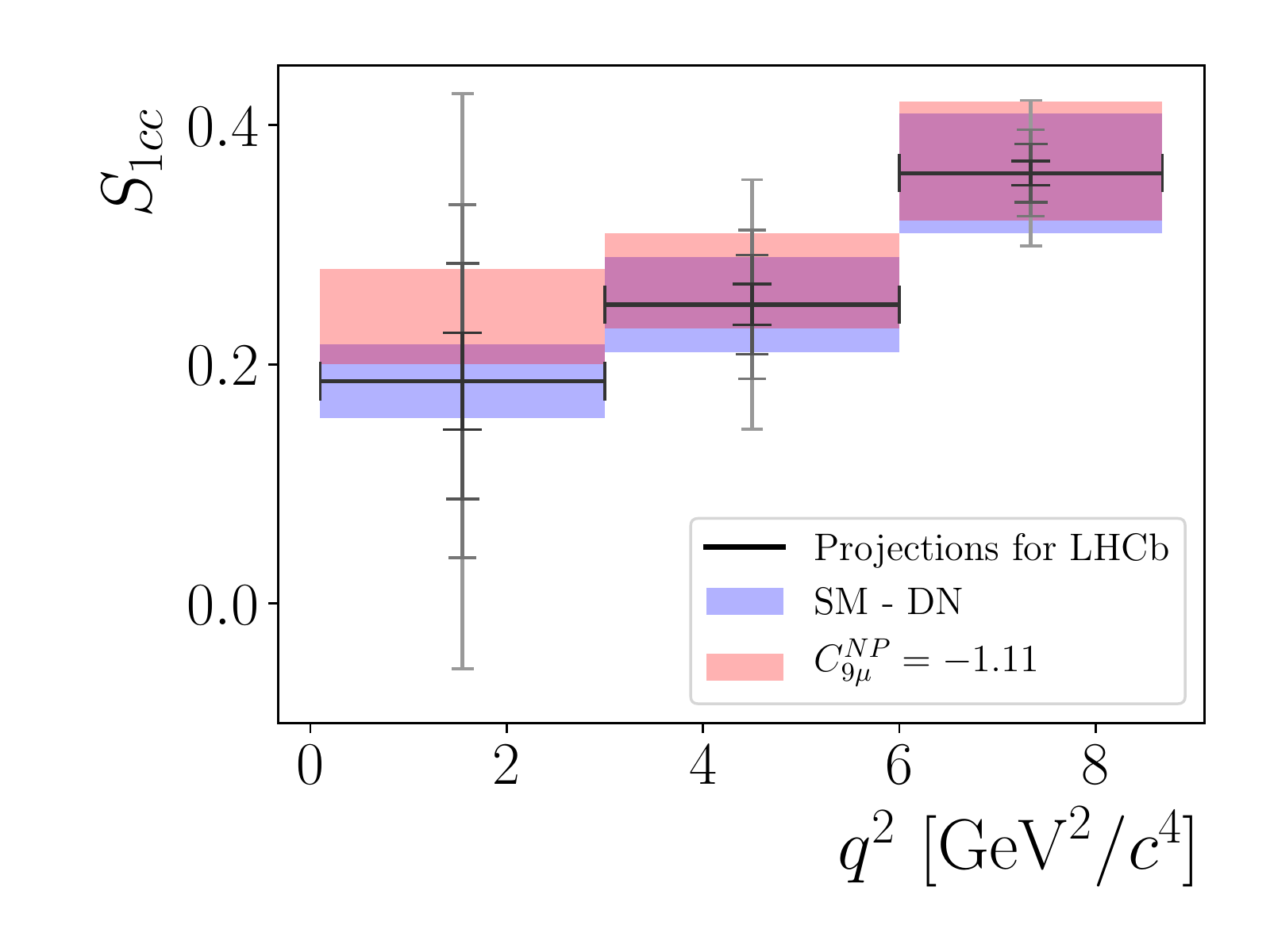}   
    \includegraphics[width=0.49\textwidth]{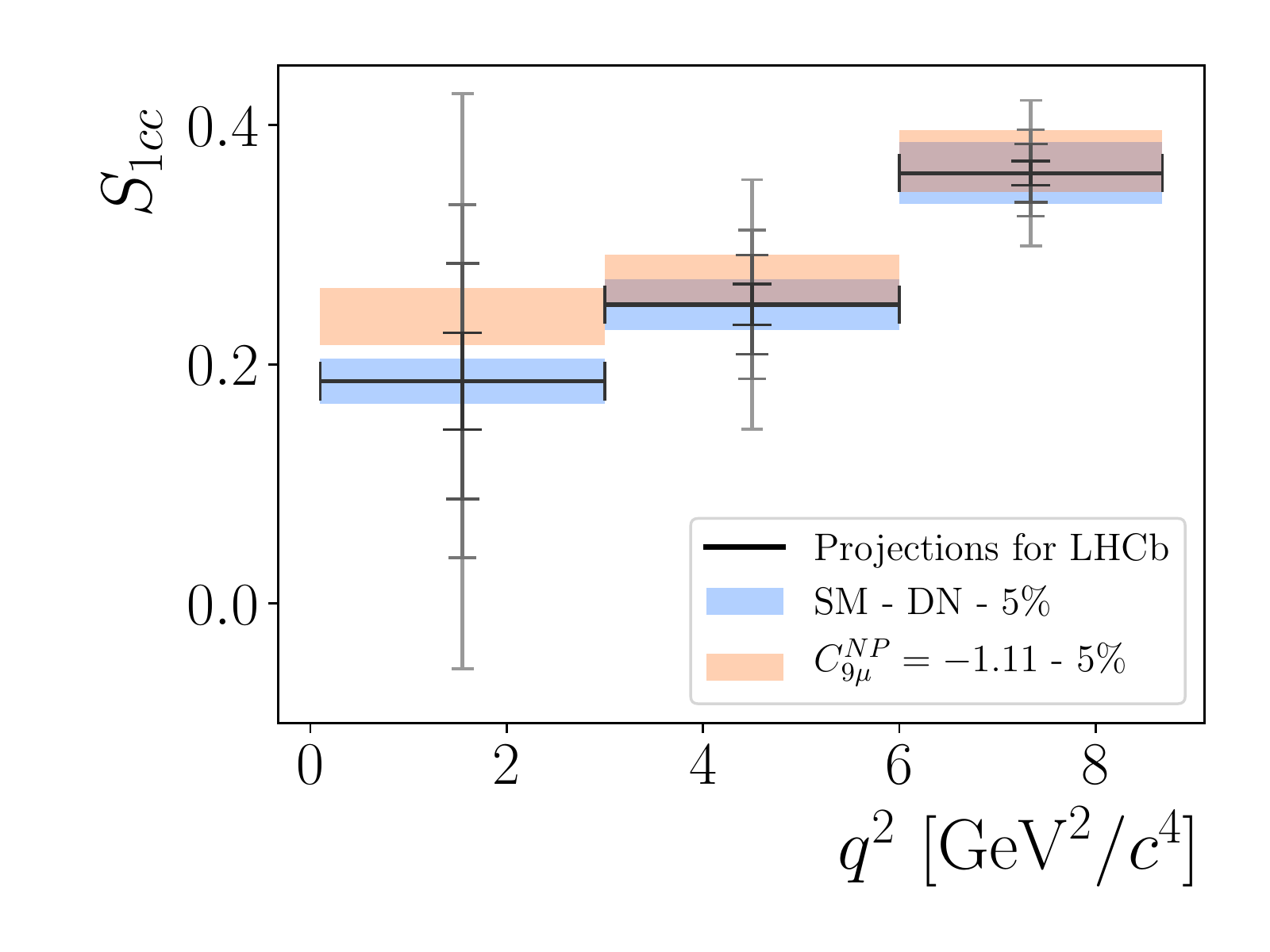}
    \caption{Theory predictions for $S_{1cc}$ in the considered \qsq bins in the SM (blue area) and the NP scenario with $C_{9\mu}^{\rm NP}= -1.11$ (red area) from Ref.~\cite{Descotes-Genon:2019dbw}, using the nominal theory uncertainties (left) 
    and improved ones with 5\% uncertainties on the form factors (right). In both cases an uncertainty of 10\% is included on \C{9} to account for $c\bar c$ contributions. 
    The expected LHCb sensitivity with the 
    full Run 2, Run 3, Run 4 and Upgrade 2 samples is shown by the grey-scale markers in increasing sensitivity.}
    \label{fig:sensitivity_S1cc}
\end{figure}

The sensitivity for different accumulated statistics is compared to the theoretical predictions in 
Figs.~\ref{fig:sensitivity_dG}, \ref{fig:sensitivity_AFB} and \ref{fig:sensitivity_S1cc} for the \dG, \AFB and \Sonecc observables, 
respectively, for the three narrow \qsq bins. 
The values for all the bins are also reported in Table~\ref{tab:exp_uncert}, where one can observe 
the experimental improvement in the broader $[0.1, 6]\gevgevcccc$ bin.
With the conservative theoretical uncertainties 
used in Ref.~\cite{Descotes-Genon:2019dbw} the decay width provides little sensitivity to separate the SM
from the NP scenario studied. However, with improved uncertainties on the form factors at the level of $5\%$,
one can disentangle with precision the two scenarios in the \qsq bins $[1, 6]$ and $[6, 8.68]\gevgevcccc$.
For a better visualisation, the relative values with respect to the SM prediction in each bin are shown in the bottom plots of Fig.~\ref{fig:sensitivity_dG} for all the \qsq bins considered.
In this case, the bin number, which follows the order presented in Tables~\ref{tab:theory_bins},~\ref{tab:yields} and~\ref{tab:exp_uncert} is used in the x-axis.
For the angular observables, while \Sonecc exhibits a poor sensitivity to NP, \AFB is more promising. 
In the conservative scenario for theory uncertainties, one could statistically separate the SM and the studied NP
model with the data sample collected by \lhcb after Upgrade 2. If the theory uncertainties
on the form factors can be reduced down to $5\%$, a good separation is already achieved after Run 3 in the 
\qsq bins [1, 6] and [6, 8.68]\gevgevcccc. 

Potential biases arising from the usage of the simplified angular PDF in Eq.~(\ref{eq:angobssimpl})
are checked by generating a large number of pseudoexperiments with the full PDF, Eq.~(\ref{eq:angobs}), 
and fitting the observables of interest, \AFB and \Sonecc, with the simplified PDF. 
At small signal yields no effect can be observed, while  
a small bias is found, which is less than $10\%$ of the statistical uncertainty, with the events expected during Upgrade 2 of \lhcb. 
This study confirms that Eq.~(\ref{eq:angobssimpl}) is a safe approximation to apply at least until $300\invfb$ have been 
recorded by \lhcb.

\section{Conclusion}
\label{sec:conclusion}

The persistent deviations in $b\to s\mu^{+}\mu^{-}$ decays and the hints of violation of lepton-flavour universality between electrons and muons in these modes provide a strong incentive to look for confirmations using other modes with different theoretical and experimental uncertainties. In this article, we presented the prospects of angular analyses of  \LbToLstll decays, motivated in particular by recent results on lepton-flavour universality in $\Lb\to pK\ell^{+}\ell^{-}$ at LHCb.

We first recalled the theoretical framework needed to analyse the \LbToLstll transition, separating short and long-distance contributions. The involvement of spin-1/2 and spin-3/2 states yields a fairly complicated differential decay rate in terms of 12 angular observables with many hadronic inputs involved, but the heavy-quark limit provides significant simplifications that are supported by the theoretical estimates currently available for the form factors. The angular distribution then factorises into the product of two terms, i.e. a trivial dependence on the angle describing the hadronic final state and a nontrivial dependence on the angle describing the leptonic final state. The three observables can be reexpressed as the branching ratio, the lepton forward-backward asymmetry and a third angular observable $S_{1cc}$. The first two observables present some sensitivity to NP contributions to the short-distance Wilson coefficient $C_{9}$ for $b\to s\mu^{+}\mu^{-}$ transitions.

Using the expected yield from the data to be
collected at the LHCb experiment in a near future, sensitivity studies were presented to determine the experimental precision on angular observables related to the lepton distribution and their potential to identify New Physics. 
We studied the impact of acceptance effects on the extraction of these angular observables using published LHCb data together with the fast simulation software {\tt RapidSim}.
The lepton forward-backward asymmetry \AFB seems particularly promising: depending on the progress made in reducing the uncertainties on the theory predictions, at some point between Run 3 and Upgrade 2, one could use this observable 
at low dilepton invariant mass
to distinguish between the SM and a scenario with NP contributions to $C_{9}$ supported by the current $b\to s\ell^{+}\ell^{-}$ data. We checked that our conclusions were not biased by the significant simplifications of the angular distribution that we proposed based on the heavy-quark limit and supported by phenomenological estimates.

We hope that our study will motivate further theoretical and experimental studies of the  \LbToLstll transitions, which could then constitute a new stepping stone to definite conclusions on the presence of New Physics in $b\to s\ell^{+}\ell^{-}$ transitions.

\section*{Acknowledgements}
%We thank UNICAFA (UNIted Crocodile and Alligator Funding Agency) for its support during this project.
C. Marin Benito  acknowledges the support  of  the  Agence  Nationale de la Recherche through the ANR grant BACH (ANR-17-CE31-0003).

\appendix 
\clearpage
\section{Angular acceptance}
\label{appendix}

%We start by discussing the acceptances.

\begin{figure} [h]
    \centering
  \includegraphics[scale = 0.40]{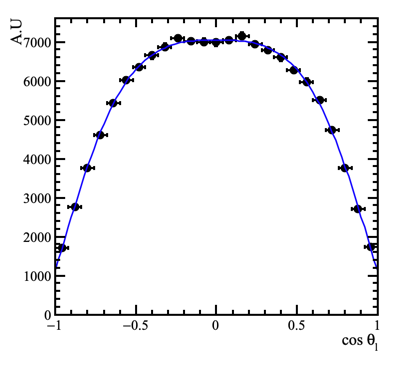}
  \includegraphics[scale = 0.40]{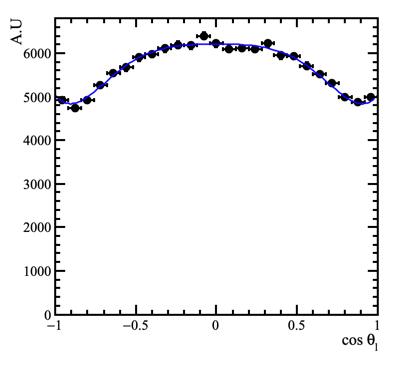}
   \includegraphics[scale = 0.40]{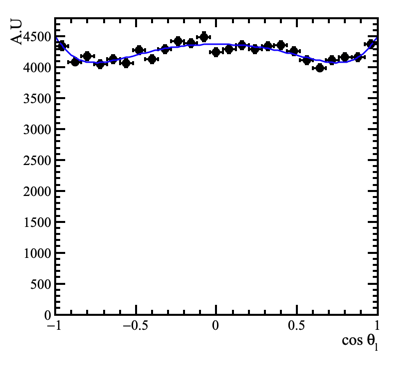}
   \includegraphics[scale = 0.40]{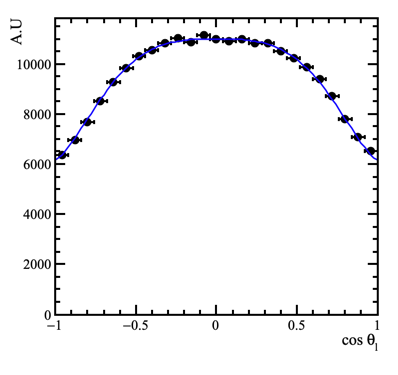}
  \caption{Acceptance shapes for $\cos\theta_\ell$ for \qsq in $[0.1,3], [3,6], [6,8.68]$, and $[1,6]$\gevgevcccc respectively.}
      \label{fig:acceptances-leptonic}
\end{figure}

The $\mu^+$ and $\mu^-$ efficiencies, introduced by the selection requirements discussed in Section~\ref{sec:sensitivity} being the same, give the acceptance shown in Figure~\ref{fig:acceptances-leptonic}, that can be modelled by an even function of Legendre polynomials:

\begin{tabular}{ll}
$\epsilon (\cos\theta_\ell)$ = &1 \\
     & + $ c_{2}/2  (3\cos^{2}\theta_\ell-1)  $ \\
     & + $ c_{4}/8  (35\cos^{4}\theta_\ell- 30 \cos^{2}\theta_\ell+3)$ \\
     & + $  c_{6}/16 (231 \cos^{6}\theta_\ell -315\cos^{4}\theta_\ell + 105 \cos^{2}\theta_\ell -5 ) $. 
\end{tabular}

\noindent
The parameters $c_i$ are fitted in simulation and used throughout the sensitivity studies. 

\newpage
\addcontentsline{toc}{section}{References}

%\setboolean{inbibliography}{true}
\bibliographystyle{LHCb}
\bibliography{main}

\end{document}